# A KNOWLEDGE BASED SYSTEM APPROACH IN SECURING DISTRIBUTED WIRELESS SENSOR NETWORKS

Daniel-Ioan Curiac*, Florin Dragan*, Ovidiu Banias* and Daniel Iercan*

* Automatics and Applied Informatics Department, "Politehnica" University Timisoara, Bd.V. Parvan Nr.2, 300223 Timisoara, Romania
Phone: (+40) 256-40-3251, Fax: (+40)-256-40-3214, E-Mail: daniel.curiac@aut.upt.ro, WWW: http://www.aut.upt.ro/~curiac

*Abstract* – *Sensor networks technologies had proved their great practicability in the real world, being just a matter of time until this kind of networks will be standardized and used in the field. This paper presents a new approach to secure the transmission of information in sensor networks and is based on a combined hardware-software architecture using three components: a) a mechanism to provide sensor authentication and secret key distribution; b) AES symmetrical encryption algorithm with predistributed keys; and c) an attack detection stratagem using an expert system based on the prediction of the values provided by the sensors, followed by reducing the sensor trust coefficient.*

*Keywords: sensor network, authentication, symmetric cryptography, knowledge based system.*

## I. INTRODUCTION

Wireless sensor networks are one of the most quickly developing domains that are using wireless communications, providing a seamless connection between the physical world and the global information infrastructure. This connection is established via a dense deployment of inexpensive, tiny sensing devices sometimes in resource-limited and hostile environments such as disaster areas, seismic zones, ecological contamination sites, military combat zones, etc. Their deployment in such harsh environments can be dangerously disturbed by any kind of sensor malfunction or, more damaging, by malicious attacks from an adversary. In this research area, it is crucial the development of new architectures/security systems that achieve protection by encrypting the transmissions, as well as protection against physical attacks on sensors (sensor capturing) or against attacks performed due to insecure routing protocols.

The design and deployment of wireless sensor networks faces some major challenges nowadays because of severe constraints which oftentimes conflict with each other, like: a) sensor nodes have restrictive power consumption, bandwidth radio transmission, computational capacity and memory; b) sensor nodes are prone to failure or malicious attacks; and, c) the aggregated sensing information provided by the sensor network must have the following attributes: accuracy, security, availability. One important challenge that researchers have to face is information security in wireless sensor networks. Even more protocols/strategies had been proposed, none can be used as a standard and none can guaranty a very good security. The motes being deployed in a certain environment without any previous network setup should be able to communicate secure and be able to avoid intruders. For this reason cryptographic algorithms should be helpful, but unfortunately not enough for the moment. The sensor network designer should decide on more security implying high power consumption and less security with low power consumption. For example, in wireless sensor networks used for detection (fire, flood, object movement, etc.) the information validity should be no more than few seconds, enough to be propagated to the sink (the sink of a wireless sensor network is a device that reports gathered information to monitoring groups or fire an alarm). In this situation, on one hand, the level of information security is not very high, and could be used symmetric cryptographic algorithms or asymmetric algorithms with shorter key length, but on the other hand, authentication should be very important in order to minimize the possibility of an attacker to send wrong information.

After studying the most important attacks in sensor networks, we considered a security scheme based on the use of three pillars: one-time authentication of sensors done just after their deployment in the field; symmetric cryptography, with pre-distributed keys for the communications inside sensor network; a knowledge based system (KBS) developed for discovering and treating the attacks in sensor networks or the sensor malfunctions. This paper will present our approach upon such a KBS and the related security method.

## II ATTACKS IN SENSOR NETWORKS

Sensor networks due to their restrictive constraints are vulnerable to some relevant types of attacks that cannot be avoided only by cryptography: eavesdropping, traffic analysis, spoofing, selective forwarding, sinkhole attack, wormhole attack, Sybil attack, Hello flood attack and node capture attack are the most important [1].

## A. Eavesdropping

Because of the nature of the carrying medium (broadcast) in wireless sensor networks, an attacker using a high-quality receiver could simply listen to data and control traffic. The adversary would be capable to retrieve information like location of node, Message IDs, Node IDs, timestamps, application specific information, etc. Strong encryption techniques should be used to counter eavesdropping.

## B. Traffic analysis

Traffic analysis is based on intercepting and inspecting messages in order to obtain information from patterns. It can be performed even when the messages are encrypted and cannot be decrypted. For example, a boost in the number of transmitted packets between certain nodes could signal that a specific sensor has registered activity.

## C. Spoofed, altered, or replayed routing information

The most direct attack against a routing protocol is to target the routing information exchanged between nodes. By spoofing, altering, or replaying routing information, adversaries may be able to create routing loops, attract or repel network traffic, extend or shorten source routes, generate false error messages, partition the network, increase end-to-end latency, etc.

## D. Selective forwarding

Selective forwarding is a subtle attack in which malicious nodes correctly forward most of the messages and selectively, silently and intentionally drop other messages which are not propagated any further.

## E. Sinkhole attack

In a sinkhole attack, an adversary is making a malicious node to look especially attractive for the surrounding nodes, for example by claiming a short or a fast route to the destination with respect to the routing metric or by making the malicious node look like a base station. By this, the attacker can create a sphere of influence and lure nearly all the traffic from that particular area through this sinkhole node.

## F. Sybil attack

In a Sybil attack, a single node illegitimately claims multiple identities to other nodes in the network.

## G. Wormholes

In the wormhole attack, an adversary makes a tunnel between two malicious nodes: one node records packets in one location and sends this information through a low-latency link to another malicious node far away, which replays it locally. This tunneling enables the tunneled packet to arrive sooner by using fewer hops than if the packet traversed a normal path.

## H. Hello flood atack

In order to announce themselves to their neighbors, in many protocols, a HELLO message is required to be broadcasted. By this, an attacker using a powerful radio transmitter sends such HELLO messages to the entire network convincing a large majority of nodes to consider the malicious node as neighbor even if it is probably out of range. The network is left in a state of confusion.

## I. Acknowledgement spoofing

If a protocol uses implicit or explicit link-layer acknowledgements, these acknowledgements can be forged, so that other nodes believe a weak link to be strong or disabled nodes alive.

## J. Node capture

Probably the biggest threat for a wireless sensor networks is node capturing attack where an adversary gains full control over sensor nodes through direct physical access [2]. This type of attack is fundamentally different from the attacks already presented because it doesn't rely on security holes in protocols, broadcasting, operating systems, etc. It is based on the geographic deployment of the sensor nodes in the field. Realistically, we cannot expect to control access to hundreds of nodes spread over several kilometers and, by this, we make a node capturing attack being very possible. An attacker can damage or replace sensor and computation hardware or extract sensitive material such as cryptographic keys to gain unrestricted access to higher levels of communication. Moreover, all sensors are usually assumed to run the same software, in particular, the same operating system. By finding an appropriate bug in the sensor network, through reverse engineering techniques applied to captured sensor, allows the adversary to control the entire sensor network.

Our proposed security architecture was developed to provide efficient countermeasures for the types of attacks presented bellow. Moreover this architecture was meant to be a realistic approach by taking into considerations the severe constraints that characterize sensor nodes.

## III. SECURITY ARCHITECTURE FOR SENSOR NETWORKS

### A. Sensor Network Model

a) We assume a static sensor network, i.e., sensor nodes are not mobile. Each sensor node knows its own location [3] even if they were deployed via aerial scattering or by physical installation. If not, the nodes can obtain their own location through location process described in [4].

   b) The sensor nodes are similar in their computational and communication capabilities and power resources to the current generation sensor nodes, e.g. the Berkeley MICA2 motes. We assume that every node has space for storing up to hundreds of bytes of keying materials.

   c) The base station also called access point, acting as a controller and as a key server, is assumed to be a laptop class device and supplied with long-lasting power. We also, assume that the base station will not be compromised.

   d) We rely on wireless cellular network (WCN) architecture [5]. In this architecture, a number of base stations are already deployed within the field. Each base station forms a cell around itself that covers part of the area. Mobile wireless nodes and other appliances can communicate wirelessly, as long as they are at least within the area covered by one cell.

   e) Also, it it possible to extend our methodology to a SENMA (SEnsor Network with Mobile Access) architecture that was proposed in [6] for large scale sensor networks. The main difference related to the cellular network architecture is that base stations are considered to be mobile, so each cell has varying boundaries which implies that mobile wireless nodes and other appliances can communicate wirelessly, as long as they are at least within the area covered by the range of the mobile access point.

   f) We are also assuming that an event can be detected by multiple sensors nodes situated in the surrounding area, so we rely on the redundancy of sensor networks.

The two types of architectures presented bellow (WCN and SENMA) have important properties that will be considered for developing a secure sensor network: nodes talk directly to base stations; no node-to-node communications; no multi-hop data transfer.; sensor synchronism is not necessary; sensor do not listen, only transmit and only when polled for; complicated protocols avoided; reliability of individual sensors much less critical; system reconfiguration for mobile nodes not necessary.

*B. First Pillar: Sensor Authentication and Key Predistribution*

Sensor authentication and key establishment in sensor networks are delicate issues since asymmetric key cryptography is inappropriate for use in resource constrained sensors. In order to solve this problem we chose the random-pairwise keys scheme presented in [7], which perfectly preserves the secrecy of the rest of the network when any node is captured, and also enables node-to-node authentication and quorum-based revocation.

*C. Second Pillar: Symmetric Encryption*

We surveyed some block ciphers to find one that is well-suited for sensor networks. Even if RC5 and Skipjack algorithms seemed to be most appropriate for software implementation on embedded microcontrollers [8], we found that AES can also be implemented efficiently, with performance not much worse than RC5 and Skipjack [9]. We consider that AES is a perfectly suitable block cipher for sensor networks although it has a small disadvantage in the fact that its block length is longer.

*D. Third Pillar: Detecting Attacks in Sensor Networks By Using a Knowledge Based System*

Detecting anomalies and intruders in sensor networks is very important. After detection, the sensor network can take decisions to investigate, find and remove malicious nodes if possible. In order to solve the problem of attacks detection, we rely on one important natural feature of sensor networks that is inherent redundancy. New approaches for ensuring security and power savings in sensor networks are based on this characteristic. It is known that redundancy in sensor networks can provide higher monitoring quality [10][11], by exploiting the adjacent nodes to discern the rightness of local data. These highly localized results can be aggregated by methods such as [12][13] to provide higher data reliability to requesting applications such as event/target detection [12][14][15]. Here, we will take a step forward connected to this approach: we will use redundancy as a feature that can bring a higher level of information security in sensor network.

There are two possible approaches: hardware redundancy and analytical redundancy. Hardware redundancy implicates the use of supplementary sensors (in normal circumstances they are already deployed in the field due to the necessity of covering the area in case of malfunctioning of some sensor nodes) and selection of data that appears similarly on the majority of sensors. Analytical redundancy is done through a process of comparison between the actual sensor value and the expected/estimated sensor value. This approach is based on a mathematical model that can predict the value of one sensor by taking into consideration the past and present values of neighboring sensors. The computational cost of this approach can become prohibitive as the number of sensors and model complexity is increased, but it can be done in our methodology at base station level (laptop class device) where all requirements are satisfied.

Our strategy uses the analytical redundancy and relies on a knowledge based system (KBS) placed in base stations (Fig.1). The principle is the following: a malicious sensor node that will try to enter false information into the sensor network will be identified by comparing its output value $x$ with the value predicted using past/present values provided by contiguous sensors $\hat{x}$. Taken into consideration a specific node denoted with A (Fig.1), this process is done in the following steps:

a) Associate a trust factor $b$ with every sensor node. In the beginning all non-marginal sensors have the same high

value of trust factor except peripheral sensors that must have a lower trust factor due to the fact that these sensors have a greater probability to be a subject of a node capture attack. For the specified sensor node A, we will have a trust factor denoted by $b_A$.

b) Estimate the future value $\hat{x}_A(t)$ provided by sensor node A, using the past/present values of adjacent sensors and the trust factor of each sensor; For the sensor denoted with A, we can write the following equation: $\hat{x}_A(t) = f(X_{A,adj}(t-1),..., X_{A,adj}(t-n), B_A)$ (1) where $X_{A,adj}(t-i) = (x_{A,adj1}(t-i),..., x_{A,adjm}(t-i))^T$ (2) is a vector that contains the values provided by all the $m$ adjacent sensors for the sensor A at the $(t-i)$ moment of time; $B_A = (b_{A,adj1},..., b_{A,adjm})^T$ (3) is a vector that contains the trust factors for each m adjacent nodes for node A; and $n$ is the estimator's order.

c) Compare the present value $x_A(t)$ of sensor node with the estimated value $\hat{x}_A(t)$ by computing the error $e_A(t) = x_A(t) - \hat{x}_A(t)$ (4);

d) Increase/decrease the trust factor $b_A$ by using a function g that can be either linear or non-linear: $b_A(t) = b_A(t-1) - g(e_A(t))$ (5)

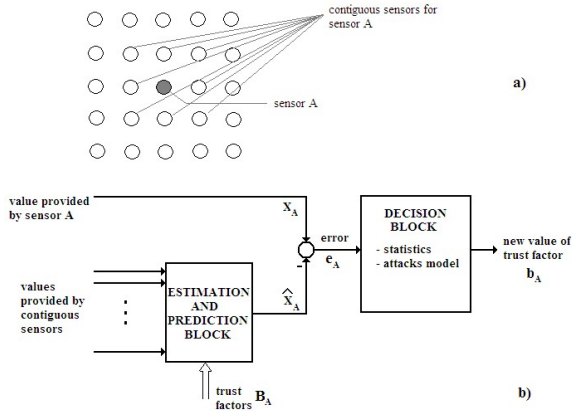

Fig. 1 : Knowledge Based System Structure – The Third Pillar of Our Architecture

The structure of such a KBS is depicted in Fig. 1 and contains two important blocks:

- Estimation and Prediction Block: this block provide the estimate $\hat{x}_A$ following (1) equation and is able to memorize the past values provided by adjacent sensors and the related trust factors. The implementation of (1) can be done using conventional prediction algorithms or predictors based on neural networks or fuzzy logic.
- Decision Block: here, based on a priori information (statistics, attack's model), the trust factor $b_A$ is modified using (5), and, in particular circumstances, some alarm signals can be transmitted to a higher hierarchical level.

## IV. CONCLUSION

Researching security for sensor networks should be at least as important as self-organization, aggregation and communication protocols. Without security the sensor networks cannot be trusted and won't have practical implementation in real world. The goal of our research was to provide security architecture to combat malicious attacks. This architecture is based on three pillars: one-time authentication done immediately after sensor deployment in the field, AES symmetric key encryption of all transmissions between sensors and base station and a KBS developed for prevention, detection and taken decisions in case of malicious attacks.